\documentclass[12pt]{article}
\usepackage{graphicx}
\usepackage[bookmarksnumbered,colorlinks,plainpages]{hyperref}

\begin{document}

\title{{\tiny Invited talk at the $12^{th}$ UN/ESA workshop on basic space science, Beijing, China, 24-28 May 2004,} \\
{\tiny to be published in Astrophysics and Space Science, 2004}\\
Maximum entropy change and least action principle for nonequilibrium
systems}

\author{Q.A. Wang\\
{\it Institut Sup\'erieur des Mat\'eriaux et M\'ecaniques Avanc\'es}, \\
{\it 44, Avenue F.A. Bartholdi, 72000 Le Mans, France}}

\date{}

\maketitle

\begin{abstract}
A path information is defined in connection with different possible paths of
irregular dynamic systems moving in its phase space between two points. On the basis
of the assumption that the paths are physically differentiated by their actions, we
show that the maximum path information leads to a path probability distribution in
exponentials of action. This means that the most probable paths are just the paths
of least action. This distribution naturally leads to important laws of normal
diffusion. A conclusion of this work is that, for probabilistic mechanics or
irregular dynamics, the principle of maximization of path information is equivalent
to the least action principle for regular dynamics.

We also show that an average path information between the initial phase volume and
the final phase volume can be related to the entropy change defined with natural
invariant measure of dynamic system. Hence the principles of least action and
maximum path information suggest the maximum entropy change. This result is used for
some chaotic systems evolving in fractal phase space in order to derive their
invariant measures.

\end{abstract}

\section{Introduction}
This work is an investigation, in connection with the concepts of information,
entropy and least action principle of Maupertuis, of the thermostatistics of dynamic
system arbitrarily away from equilibrium. It is hoped that the results may have some
connections with quantum gravity theories, quantum cosmology and quantum
chromodynamics in particle physics because, firstly, most, if not all, of the
systems treated in these fields are in nonequilibrium state and, secondly, the
statistical concepts of probability, information, entropy and least (stationary)
action solutions (instantons) to Feynman path integrals are so essential and widely
used in these disciplines\cite{Hawking}. An important character of the path integral
methods is that the probability of certain evolution is determined by an exponential
functional of action which is $postulated$ for the first time by
Feynman\cite{Feynman}. The reader will find that this is a natural consequence of
the principle of maximum path information combined with action of the trajectories
of nonequilibrium evolution.

Although the theoretical formalism of equilibrium statistics is well founded on the
physical laws with the help of the algorithm of maximum entropy\cite{Jaynes,
Tribus}, the present statistical methodology for nonequilibrium systems seems to
some extent uncertain\cite{Ruelle}. For some nonequilibrium systems, the extension
of the concept of Boltzmann and Gibbs entropies are in
progress\cite{Ruelle,Cohen,Lebowitz}. For stationary state sufficiently close to
equilibrium, the principle of minimum entropy production is
proposed\cite{Prigogine}. For certain systems far from equilibrium, e.g., in the
context of the climate of earth, there is the method of maximum entropy production
(or minimum entropy exchange)\cite{Paltridge,Lenton}. Here the entropy is always in
the sense of Clausius, so the entropy production $dS_i\geq 0$ according to the
second law. Recently, we have witnessed the development of nonextensive statistics
theories based on the Tsallis entropy which has been maximized for nonequilibrium
systems\cite{Tsallis,Wang02c} in chaotic motions or on the edge of chaos according
to actual understanding. This is nothing but an extension of Jaynes principle for
nonequilibrium systems\cite{Jaynes2} to another entropy which is not necessarily an
entropy in the sense of Clausius.

This work is intended to study the behavior of irregular dynamic systems from the
point of view of information theory. It is expected that the reasoning of this work
concerning the principle of maximum information and entropy in connection with the
concept of action and the principle of Maupertuis, can be helpful for understanding
the physics of dynamic processes. In the second part of this paper, we specify some
definitions and assumptions used in this work. In the third part, we describe a path
information\cite{Wang04x} defined in connection with different possible paths of
nonequilibrium systems moving in its phase space between two points (cells). On the
basis of the assumption that the paths are physically differentiated by their
actions, we show that the maximum path information yields an exponential path
probability distribution depending on action which implies the most probable paths
are just the paths of least action\footnote{During the preparation of this
manuscript, Dr. Touchette wrote me that this connection was known in the field of
random perturbation of dynamic systems and large deviation. The possible references
are: M.I. Freidlin, A.D.Wentzell, Random Perturbations of Dynamic Systems,
Springer-Verlag, 1983; and Y. Oono, Large deviation and statistical physics,
Progress of Theoretical Physics Supplement 99, 165-205, 1989. I am most indebted to
him for pointing out that.}. Finally, this {\it least action distribution} is
differentiated with time and position in order to derive, in a quite general manner
without additional assumptions, the Fokker-Planck equation, the Fick's laws of
normal diffusion, the Ohm's law of electrical conduction and the Fourier's law of
heat conduction.

In the forth part, an averaged path information is related to the difference of
entropy between two phase volumes. The entropy is defined with the natural invariant
measure of the nonequilibrium system. This relationship suggests that, if one
maximizes path information in order to know the probability distribution of paths
for dynamic systems whose motion forms geodesic flows in phase space, the entropy
change of the dynamic process must be maximized in order to derive the invariant
measures of the process.

Finally, the above result is applied to some chaotic systems evolving in fractal
phase space for which a relative entropy change is given by
$R=\sum_i\mu_i-\sum_i\mu_i^q$, where $q$ is a positive real parameter characterizing
the geometrical features of the phase space and $\mu_i$ a natural invariant measure.
Its maximization (or extremization) with appropriate constraints can yield several
power law invariant measures.

\section{Assumptions and definitions}

Let us begin by some assumptions and definitions which specify the range of validity
of this work.

\begin{enumerate}

\item Phase space and partition. As usual the phase space $\Gamma$ of a
thermodynamic system is defined such that a physical state of the system is
represented by a point in that space. For a $N$-body system moving in three
dimensional ordinary configuration space, the $\Gamma$-space is of $6N$ dimension
($3N$ positions and $3N$ momenta) if it can be smoothly occupied. The position of
the state point at $t=0$ of a movement is called initial condition of that movement.
We suppose that a phase volume $\Omega$ accessible to the system can be partitioned
into $v$ cells of volume $s_i$ with $i=1,2,...v$ in such a way that $s_i\cap
s_j=\emptyset$ ($i\neq j$) and $\cup_{i=1}^vs_i=\Omega$. A state of the system can
be represented by a sufficiently small phase cell in coarse graining way. The
movement of a dynamic system can be represented by its trajectories (in the sense of
classical mechanics) in $\Gamma$ space.

\item The natural invariant measure\cite{Beck} $\mu_i$ (probability distribution for a
nonequilibrium system to visit different phase cells of given partitions) is defined
for the cell $i$ as follows :
\begin{eqnarray}                                            \label{+1}
\mu_i=\int_{s_i}dx \rho(x)
\end{eqnarray}
where $\rho(x)$ is the probability density and $x$ represents the phase position.
For dynamic systems viewed as geodesic flow in phase space, the phase trajectories
is uniformly distributed in the accessible phase volume\cite{Arnold}. In this case,
$\rho(x)$ can be constant over all the phase space and $\mu_i$ becomes proportional
to $s_i$, a phase cell of a given partition. This property will be useful for
calculating entropy change of a system moving in fractal phase space (see below).

\item The ergodicity for nonequilibrium system is described by
\begin{eqnarray}                                            \label{+2}
\overline{Q}=\lim_{T\to\infty}\frac{1}{T}\int_0^T Q(t)dt
&=&\lim_{v\to\infty}\sum_{i=1}^{v}\mu_iQ_i.
\end{eqnarray}
which means that the average over time is equal to the ensemble average over the
occupied phase space volume, where $T$ is the period of the evolution, $Q(t)$ is the
value of a quantity $Q$ at time $t$ and $Q_i$ is the value of $Q$ in the cell $i$. In
this way, the evolution of information over time can be estimated by the time
evolution of the phase space volume occupied by the statistical ensemble of identical
systems under consideration.

\item The information we address in this work is our ignorance about the system
under consideration associated with some uncertainty or probability distribution.
According to Shannon, this information can be measured by the formula
\begin{eqnarray}                                            \label{++2}
H=-\sum_{i=1}^vp_i\ln p_i
\end{eqnarray}
with respect to certain probability distribution $p_i$ and the index $i$ is summed
over all the possible situations or events.

\item In this work, entropy $S$ of a nonequilibrium system is defined with the natural
invariant measure $\mu_i$ either by Eq.(\ref{++2}) or by other formula, i.e.,
\begin{eqnarray}                                            \label{++2+}
S=-\sum_{i=1}^v\mu_i\ln \mu_i=-\int_\Omega\rho(x)\ln\rho(x)dx
\end{eqnarray}
or, more generally, $S=\int_\Omega \sigma[\rho(x)]dx$, where $\Omega$ is the total
occupied phase volume and $\sigma$ is the entropy density. In the special case of
Eq.(\ref{++2+}), $\sigma[\rho(x)]=\rho(x)\ln\rho(x)$.

\item According to the ergodic assumption mentioned above, the entropy change
$\Delta I(T)$ of a dynamic system during a long time process can be estimated by the
ensemble average of the entropy change during the scale refinement,

\begin{eqnarray}                                            \label{+3}
\Delta I(T) &=& \lim_{T\to\infty}\frac{1}{T}\int_0^T \delta i(t)dt \\
\nonumber & \Longrightarrow &
\lim_{v_T\to\infty}\sum_{i=1}^{v_T}\mu_i\int_{s_i}\sigma_ids
-\sum_{j=1}^{v_0}\mu_j\int_{s_j}\sigma_jds  \\ \nonumber
&=&\lim_{v_k\to\infty}\sum_{i=1}^{v_k}\mu_i\int_{s_i}\sigma_ids
-\sum_{j=1}^{v_0}\mu_j\int_{s_j}\sigma_jds
\end{eqnarray}

where $\Delta I(T)$ is the average entropy change during the time $T$, $\delta i(t)$
is the entropy change at time $t$, $\sigma_i$ is the density of information on the
phase cell $i$ of volume $s_i$, $v_T$ and $v_0$ are the total numbers of the cells of
phase space accessible to the system and visited by the trajectories at time $t=T$ and
$t=0$, respectively. According to our assumption, the $v_T$ cells visited by the
system form a geometry which can be reproduced by the $v_k$ cells yielded from the
$v_0$ cells by certain map (scale refinement) of $k$ iterations. We may put $v_k=v_T$
when $k$ and $T$ are large.

\end{enumerate}

\section{About a path information}
\subsection{Uncertainties of irregular dynamics}
In a previous work\cite{Wang04x}, we defined a path information for irregular
dynamic systems moving in the $\Gamma$-space between two points, $a$ and $b$, which
are in two elementary cells of a given partition of the phase space. It is known
that if the motion of the system is regular, there will be only a fine bundle of
paths which track each other between the initial and the final cells and minimize
action according to the principle of least action\cite{Arnold}. However, if the
dynamics is irregular due to large number of degrees of freedom or strong
sensitivity to initial conditions, we can have the following two dynamic
uncertainties:

\begin{enumerate}

\item Between any two fixed cells $a$ and $b$, there may be different possible paths
(labelled by $k$=1,2,...w) each having a probability $p_k(b|a)$ to be followed by
the system. This is the uncertainty we studied in \cite{Wang04x}.

\item In a fixed period of time, we can observe different possible paths leaving the
cell $a$ and leading to different final cells $b$ in a final phase volume $B$, each
having a probability to be followed by the system.
\end{enumerate}
In this section, the discussions will be made with the first uncertainty. By
definition, the path probability distribution $p_k(b|a)$ is a transition probability
from state $a$ to state $b$ via path $k$ for all systems between these two states.
We have $\sum_{k=1}^wp_k(b|a)=1$. The dynamic uncertainty associated with $p_k(b|a)$
is measured by the Shannon information :
\begin{eqnarray}                                            \label{c1}
H(a,b)=-\sum_{k=1}^wp_k(b|a)\ln p_k(b|a).
\end{eqnarray}
$H(a,b)$ is a {\it path information}, i.e., the missing information necessary for
predicting which path a system takes from $a$ to $b$.

\subsection{Least action distribution}
The path probability distribution $p_k(b|a)$ due to dynamic irregularity can be
studied in connection with information theory and action integral on the basis of
the assumption that {\it the different paths are uniquely differentiated by their
actions}. For classical mechanical systems, action is given by
\begin{eqnarray}                                            \label{c5}
A_{ab}(k)=\int_{t_{ab}(k)}L_k(t)dt
\end{eqnarray}
where $L_k(t)$ is the Lagrangian of the system at time $t$ along the path $k$. For
other systems, the action may be given by different calculations, e.g., Yang-Mills
action and Euclidean action for quantum field theory and quantum
cosmology\cite{Hawking}. The average action between state $a$ and state $b$ is given
by
\begin{eqnarray}                                            \label{xc5}
A_{ab}=\sum_{k=1}^wp_k(b|a)A_{ab}(k).
\end{eqnarray}

The maximization of $H(a,b)$ under the constraints associated with the normalization
of $p_k(b|a)$ and the average action leads to
\begin{eqnarray}                                            \label{c6}
p_k(b|a)=\frac{1}{Q}\exp[-\eta A_{ab}(k)]
\end{eqnarray}
Putting this distribution into $H(a,b)$ of Eq.(\ref{c1}), we get
\begin{eqnarray}                                            \label{c7}
H(a,b)=\ln Q+\eta A_{ab}=\ln Q-\eta \frac{\partial}{\partial\eta}\ln Q
\end{eqnarray}
where $Q$ is given by $Q=\sum_{k=1}^w\exp[-\eta A_{ab}(k)]$ and
$A_{ab}=-\frac{\partial}{\partial\eta}\ln Q$.

It is proved that\cite{Wang04x} the distribution Eq.(\ref{c6}) is stable with
respect to the fluctuation of action, and that Eq.(\ref{c6}) is a least (stationary)
action distribution, i.e., the most probable paths are just the paths of least
action. We have indeed $\delta p_k(b|a)=-\eta p_k(b|a)\delta A_{ab}(k)=0$, so that
$\delta A_{ab}(k)=0$ leading to Euler-Lagrange equation $\frac{\partial}{\partial
t}\frac{\partial L_{k}(t)}{\partial \dot{x}}-\frac{\partial L_{k}(t)}{\partial
x}=0$\cite{Arnold} and to Newton's second law which are satisfied by the most
probable paths. For stochastic process like Brownian motion\cite{Kubo}, it can be
proved\cite{Wang04x} that the Lagrange multiplier $\eta$ is positive and given by
the inverse of the diffusion coefficient $\eta=\frac{1}{2mD}$. So the action of
Brownian particles has a minimum along the most probable paths.

The physical content of $\eta$ can be made clearer if we use the general
relationship $D=\frac{l^2}{2\tau}$ for Brownian motion leading to
\begin{eqnarray}                                            \label{xx9x}
\eta=\frac{\tau}{ml^2},
\end{eqnarray}
where $l$ is the mean free path and $\tau$ the mean free time of the Brownian
particles. If we still consider detailed balance\cite{Kubo}, we get $D=\mu k_BT$
where $\mu$ is the mobility of the diffusing particles, $k_B$ the Boltzmann constant
and $T$ the temperature. This means
\begin{eqnarray}                                            \label{xxx9x}
\eta=\frac{1}{2m\mu K_BT}=\frac{\gamma}{2K_BT}
\end{eqnarray}
where $m\gamma=\frac{1}{\mu}$ is the friction constant of the particles in the
diffusion mixture. Eq.(\ref{c6}) can now be written as
\begin{eqnarray}                                            \label{cc6}
p_k(b|a)=\frac{1}{Q}\exp[-\frac{\gamma A_{ab}(k)}{2K_BT} ]
\end{eqnarray}

\subsection{A calculation of transition probability}
Now let us look at a particle of mass $m$ diffusing along a given path from $a$ to a
cell $b$ of its $\mu$-space (one particle phase space). The path is cut into $N$
infinitesimally small segments each having a spatial length $\Delta x_i=x_i-x_{i-1}$
with $i=1 ...N$ ($x_0=x_a$ and $x_N=x_b$). $t=t_i-t_{i-1}$ is the time interval spent
by the system on every segment. The Lagrangian on the segment $i$ is given by
\begin{eqnarray}                                            \label{x9c}
L(x,\dot{x},t)=\frac{m(x_i-x_{i-1})^2}{2(t_i-t_{i-1})^2} -\left(\frac{\partial
U}{\partial x}\right)_i\frac{(x_i-x_{i-1})}{2}-U(x_{i-1})
\end{eqnarray}
where the first term on the right hand side is the kinetic energy of the particle,
the second is the average increment of its potential energy and the third is its
potential energy at the point $x_{i-1}$. The action on the segment $i$ is just
\begin{eqnarray}                                            \label{xx9c}
A_i=\frac{m(\Delta x_i)^2}{2t} +F_i\frac{\Delta x_i}{2}t-U(x_{i-1})t,
\end{eqnarray}
where $F_i=-\left(\frac{\partial U}{\partial x}\right)_i$ is the force on the
segment $i$. According to Eq.(\ref{c6}), the transition probability $p_{i/i-1}$ from
$x_{i-1}$ to $x_i$ on the path $k$ is given by
\begin{eqnarray}                                            \label{xxxc9}
p_{i/i-1} &=& \frac{1}{Z_i} \exp\left(-\eta\left[\frac{m}{2t}\Delta x_i^2
+F_i\frac{t}{2}\Delta x_i\right]_{k}\right)
\end{eqnarray}
where $Z_i$ is calculated as follows
\begin{eqnarray}                                            \label{xxx9}
Z_i&=&\int_{-\infty}^\infty dx_i\exp\left(-\eta\left[\frac{m}{2t}\Delta x_i^2
+F_i\frac{t}{2}\Delta x_i\right]_{k}\right)\\ \nonumber &=&
\exp\left[F_i^2\frac{\eta t^3}{8m}\right]\sqrt{\frac{2\pi t}{m\eta}}.
\end{eqnarray}
The potential energy at the point $x_{i-1}$ disappears in the expression of
$p_{i/i-1}$ because it does not depend on $x_i$.

The total action is given by
\begin{eqnarray}                                            \label{cc9}
A_{ab}(k)=\sum_{i=1}^NA_i=\sum_{i=1}^N\left[\frac{m(\Delta x_i)^2}{2t}
+F_i\frac{t}{2}\Delta x_i-U(x_{i-1})t\right]_{k}
\end{eqnarray}
so the transition probability from $a$ to $b$ via the path $k_b$ is the following:
\begin{eqnarray}                                            \label{c9}
p_k(b|a) &=& \frac{1}{Z} \exp\left[-\eta\sum_{i=1}^N\left[\frac{m(\Delta x_i)^2}{2t}
+F_i\frac{t}{2}\Delta x_i\right]_{k}\right] \\ \nonumber &=& \prod_{i=1}^N p_{i/i-1}
\end{eqnarray}
where
\begin{eqnarray}                                            \label{c9x}
Z= \prod_{i=1}^NZ_i=\prod_{i=1}^N\exp\left[F_i^2\frac{\eta
t^3}{8m}\right]\sqrt{\frac{2\pi t}{m\eta}}.
\end{eqnarray}

\subsection{A derivation of diffusion laws}

\subsubsection{Fokker-Planck equation}
The Fokker-Planck equation describes the time evolution of the probability density
function of position and velocity of a particle. This equation can be derived on the
basis of the assumptions of Brownian motion and Markovian process\cite{Kubo}. We
will show here that this equation can be derived from the distributions given by
Eq.(\ref{xxxc9}) and Eq.(\ref{c9}) without any assumptions.

The calculation of the derivatives $\frac{\partial p_{i/i-1}}{\partial t}$,
$\frac{\partial F_ip_{i/i-1}}{\partial x_i}$ and $\frac{\partial^2
p_{i/i-1}}{\partial x_i^2}$ straightforwardly leads to
\begin{eqnarray}                                            \label{xc10}
\frac{\partial p_{i/i-1}}{\partial t} = -\frac{\tau}{m}\frac{\partial
(F_ip_{i/i-1})}{\partial x_i}+D\frac{\partial^2 p_{i/i-1}}{\partial x_i^2}.
\end{eqnarray}
This is the Fokker-Planck equation, where $D$ is given by $D=\frac{1}{2m\eta}$ and
$\tau$ is the mean free time supposed to be the time interval $t$ of the particle on
each segment of its path. In view of the Eq.(\ref{c9}), it is easy to show that this
equation is also satisfied by $p_k(b|a)$ if $x_i$ is replaced by $x_b$ or $x$, the
final position. Now let $n_a$ and $n_b$ be the particle density at $a$ and $b$,
respectively. The following relationship holds
\begin{eqnarray}                                            \label{xxc10x}
n_b=\sum_{k=1}^wn_ap_k(b|a)
\end{eqnarray}
which is valid for any $n_a$. This means (let $n=n_b$ and $x=x_b$):
\begin{eqnarray}                                            \label{xc10x}
\frac{\partial n}{\partial t} = -\frac{\tau}{m}\frac{\partial (Fn)}{\partial
x}+D\frac{\partial^2 n}{\partial x^2},
\end{eqnarray}
which describes the time evolution of the particle density.

\subsubsection{Fick's laws of diffusion}
If the external force $F$ is zero, we get
\begin{eqnarray}                                            \label{c10xx}
\frac{\partial n}{\partial t} = D\frac{\partial^2 n}{\partial x^2}
\end{eqnarray}
This is the second Fick's law of diffusion. The first Fick's law
\begin{eqnarray}                                            \label{cc10xx}
J(x,t) = -D\frac{\partial n}{\partial x}
\end{eqnarray}
can be easily derived if we consider matter conservation $\frac{\partial
n(x,t)}{\partial t}= -\frac{\partial J(x,t)}{\partial x}$ where $J(x,t)$ is the flux
of the particle flow.

The diffusion constant $D$ can be related to the partition function $Z_a$ by
combining the relationship $D=\frac{1}{2m\eta}$ and Eq.(\ref{c7}), i.e.,
\begin{eqnarray}                                            \label{cx10xx}
D=\frac{1}{2m}\frac{\partial A_a}{\partial H_a}=-\frac{1}{2m}\frac{\partial^2(\ln
Z_a)}{\partial H_a\partial\eta}.
\end{eqnarray}

\subsubsection{Ohm's law of electrical conduction}
Considering the charge conservation $\frac{\partial \rho(x,t)}{\partial t}=
-\frac{\partial j(x,t)}{\partial x}$, where $\rho(x,t)=qn(x,t)$ is the charge
density, $j(x,t)=qJ(x,t)$ is the flux of electrical currant and $q$ is the charge of
the currant carriers, we have, from Eq.(\ref{xc10x}),
\begin{eqnarray}                                            \label{xc10xxx}
j=\frac{\tau}{m}F\rho - D\frac{\partial \rho}{\partial x}
\end{eqnarray}
where $F=qE$ is the electrostatic force on the carriers and $E$ is the electric
field. If the carrier density is uniform everywhere, i.e., $\frac{\partial
n}{\partial x}=0$, we get
\begin{eqnarray}                                            \label{xxc10xxx}
j=\frac{\tau}{m}qE\rho=\sigma E,
\end{eqnarray}
where $\sigma=\frac{q\rho\tau}{m}=\frac{nq^2\tau}{m}$ is the formula of electrical
conductivity widely used for metals.

\subsubsection{Fourier's law of heat conduction}
For heat conduction, let the external force $F$ be zero. We consider a crystal
idealized by a lattice of identical harmonic oscillators each having an energy
$e_k=N_\nu(x,t)h\nu$ where $h$ is the Planck constant, $\nu$ is the frequency of a
mode and $N_\nu(x,t)$ is the number of phonons of that mode situated at $x$ at time
$t$ in the intervals $x\rightarrow x+dx$ and $\nu\rightarrow \nu+d\nu$. Suppose that
there is no mass flow and other mode of energy transport in the crystal. Heat is
transported only through the phonon flow. The phonons of frequency $\nu$ diffuse in
the crystal lattice, among the lattice imperfections, impurities and other phonons
with in addition anharmonic effects\cite{Bonetto}, just like a particle of mass
$m=h\nu/c^2$ having transition probability $p_k(b|a)$. Let $n_\nu=N_\nu/dx$ be the
density of phonons which must satisfy
\begin{eqnarray}                                            \label{cx10x}
n_\nu(x_b,t)=\sum_{k}n_\nu(x_a,t)p_k(b|a)
\end{eqnarray}
and also Eq.(\ref{c10xx}).

The total energy density $\rho(x,t)$ of phonons at $x$ and time $t$ is given by
\begin{eqnarray}                                            \label{cxc10}
\rho(x,t)=\int_0^{\nu_m}h\nu n_\nu(x,t)\varrho(\nu)d\nu,
\end{eqnarray}
where $\varrho(\nu)d\nu$ is the mode number in the interval $\nu\rightarrow
\nu+d\nu$, and $\nu_m$ is the maximal frequency of the lattice vibration. This
relationship holds for any state density $\varrho(\nu)$ and frequency $\nu$. This
implies
\begin{eqnarray}                                            \label{xc10xx}
\frac{\partial \rho(x,t)}{\partial t} = D \Delta_{x} \rho(x,t)
\end{eqnarray}

On the other hand, the variation of energy density $\delta\rho(x,t)$ can be related
to temperature change $\delta T(x,t)$ by
\begin{eqnarray}                                            \label{11}
\delta\rho(x,t)=c\delta T(x,t)
\end{eqnarray}
where $c$ is the heat capacity per unit volume supposed constant everywhere in the
crystal. This leads to
\begin{eqnarray}                                            \label{c12}
\frac{\partial \rho}{\partial t} = \kappa\Delta_xT(x,t)
\end{eqnarray}
where $\kappa=Dc$ is the heat conductivity. This equation can be recast into
\begin{eqnarray}                                            \label{c12xx}
c\frac{\partial T(x,t)}{\partial t} = \kappa\Delta_xT(x,t)
\end{eqnarray}
which describes the evolution of temperature distribution due to the heat flow. When
a stationary state is reached, temperature is everywhere constant, i.e.,
$\frac{\partial T(x,t)}{\partial t}=\Delta_xT(x,t)=0$, and the temperature
distribution is given by $\nabla_x T(x)=constant$.

Now considering the energy conservation in an elementary volume between $x$ and
$x+dx$ in which we have $-\frac{\partial \rho(x,t)}{\partial t}= \nabla_x\cdot
J(x,t)$, Fourier's law of heat conduction follows
\begin{eqnarray}                                            \label{c12x}
J(x,t)=-\kappa\nabla_x T(x,t).
\end{eqnarray}

Summarizing this section, we have introduced a path information \`a la Shannon for
irregular dynamic systems and assumed that the different paths between any two fixed
points in phase space are characterized by their actions. This starting points lead
to a least action transition probability in exponentials of action which maximizes
the path information and implies that the most probable paths are just the paths
(geodesics) required by the least action principle of classical mechanics. From this
least action distribution, we can derive important physical laws for normal
diffusion without any additional assumptions about the dynamic process (Brownian and
Markovian or not). This result can be considered as a robust support to this
approach of probabilistic mechanics or statistical dynamics. In what follows, the
above method concerning only two fixed phase cells will be extended to arbitrary
final cell $b$ in order to study the variation of entropy of dynamic process from
the point of view of path information.

\section{Maximize entropy change for nonequilibrium systems}

Now the above result concerning only two fixed cells will be generalized to
arbitrary destination $b$. Let us consider an ensemble of $N$ identical chaotic
systems leaving the initial cell $a$ for some destinations. The travelling time is
specified, say, $t_{ab}=t_b-t_a$. After $t_{ab}$, all the phase points occupied by
the systems form the final phase volume $B$ partitioned into cells labelled by $b$.
We observe $N_{b}$ systems travelling along an ensemble of paths labelled also by
$b$ leading to certain cell $b$. A path probability can be defined by
$p_{b|a}=N_{b}/N$ which is normalized by $\sum_bp_{b|a}=1$. We always suppose each
ensemble of paths $b$ is characterized by the average action $A_{ab}$ over the
ensemble given by Eq.(\ref{xc5}). Then a total average action can be defined by
$A_{a}=\sum_{b}p_{b|a}A_{ab}$. The uncertainty concerning the choice of final cells
is just the second uncertainty mentioned in section 3.1. It is measured by the
following Shannon information
\begin{eqnarray}                                            \label{c1x}
H_{a}=-\sum_{b}p_{b|a}\ln p_{b|a}
\end{eqnarray}
and can be maximized just as in section 3.2 to give
\begin{eqnarray}                                            \label{c6x}
p_{b|a}=\frac{1}{Z}\exp[-\theta A_{ab}],
\end{eqnarray}
where $\theta$ is the Lagrange multiplier associated with $A_{a}$. As has been done
for the distribution of Eq.(\ref{c6}) in \cite{Wang04x}, it can be proved that this
distribution is stable with respect to the fluctuation of the average action $A_{ab}$.
It has also been proved in \cite{Wang04x} that $A_{ab}$ has a stationary (minimum if
$\eta>0$ and maximum if $\eta<0$) when the uncertainty (path information) $H(a,b)$ is
maximized. This means that the ensemble of paths with stationary $A_{ab}$ are the most
probable, i.e., $\delta A_{ab}=0\;\Longleftrightarrow\;\delta p_{b|a}=0$.

Our aim here is to derive the invariant measures $\mu_i$ which define the entropy
$S$ for dynamic systems by Eq.(\ref{++2+}) or, more generally, by certain functional
$\sigma[\rho(x)]$. Our idea is to ``complement'' the method of maximum entropy by
another method for dynamic systems. This is the method of maximum entropy change.

In what follows, we suppose that entropy changes when an irregular dynamic system
moves from the initial cell $a$ to the final cells $b$ in $B$. The initial entropy
of the systems at time $t=0$ is given by Shannon formula:
\begin{eqnarray}                                            \label{x1}
S_a=-\sum_{a}^{over A}\mu_a\ln \mu_a
\end{eqnarray}
where $\mu_a$ is the invariant measure of the systems in the initial phase volume $A$
at $t=0$ and the cell index $a$ is summed over all $A$. After a time $t_{ab}$, the
systems from the cell $a$ travelling along an ensemble $b$ of paths are found in a
cell $b$ with invariant measure $\mu_b(a)$. Supposing Markovian process, we have
\begin{eqnarray}                                            \label{x2}
\mu_b(a)=\mu_ap_{b|a}.
\end{eqnarray}
We calculate first the contribution of the cell $a$ to the final entropy of the
systems in $B$. It is given by
\begin{eqnarray}                                            \label{cx3}
S_b(a) &=& -\sum_{b}\mu_{b}(a)\ln \mu_{b}(a)\\
\nonumber &=& -\mu_a\ln \mu_a-\mu_a\sum_{b}p_{b|a}\ln p_{b|a}.
\end{eqnarray}
The total final entropy is then
\begin{eqnarray}                                            \label{x3}
S_b &=& -\sum_{a}S_b(a) \\ \nonumber &=& -\sum_a\mu_a\ln \mu_a-\sum_a\mu_a\sum_{b}p_{b|a}\ln p_{b|a} \\
\nonumber &=& S_a+\sum_a\mu_aH_{a}=S_a+\overline{H}_{a},
\end{eqnarray}
where $\overline{H}_{a}$ is the average of the path information $H_{a}$ over all the
initial phase volume $A$. So if $H_{a}$ is maximized, $\overline{H}_{a}$ should also
be maximized.

Eq.(\ref{x3}) is an essential result of this work. It states that, if $H_{a}$ as
well as $\overline{H}_{a}$ must be maximized by the least action distributions, then
the entropy change $\Delta S=S_b-S_a=\overline{H}_{b|a}$ must be also maximized for
dynamic process.

We would like to emphasize here that maximum (extremum) entropy change does not
exclude the possibility of maximum (extremum) entropy. In fact, if the entropy change
is maximized at any moment of an evolution, the entropy must also be at maximum all
the time. This can be considered as a proof of the maximum entropy method derived from
action principle for stochastic process. The choice of these two methods for a given
process of course depends on which of entropy and entropy change is available for the
probability functionals to be derived. Following section is an example of the
availability of entropy change without even knowing the exact entropy functionals.

\section{The entropy change due to fractal geometry}

Now let us use the method of extremum entropy change (an extension of maximum
entropy change) to derive invariant measure of dynamic systems. We consider a
statistical ensemble of identical dynamic systems all moving in fractal phase space.
The outcome of this movement is not specified. This may be a relaxation process
towards equilibrium or a long evolution from one nonequilibrium state to another.
For this kind of systems, an entropy change has been derived in our previous
work\cite{Wang04a} in term of invariant measures. We give here a brief description
of the method of maximum entropy change and some of its consequences.

In reference \cite{Wang04a}, we supposed that all state points in the phase space were
equiprobable and uniformly distributed in the initial condition volume. This is true
if the ensemble of studied systems forms geodesic flows on phase
manifolds\cite{Arnold}. In this case, the invariant measure $\mu_i$ of a phase cell
$i$ of a given partition is just proportional to the volume $s_i$ of the cell. We
supposed in addition the state density is scale invariant, which is in accordance with
the time independent invariant measure if we consider the ergodic assumption. In this
case, the relative entropy change from zeroth to $\lambda^{th}$ iteration of the scale
refinement (finer and finer partition of the phase space) is given by
\begin{eqnarray}                                            \label{3}
R=\sum_{i=1}^{v_\lambda}(\mu_i-\mu_i^q)
\end{eqnarray}
where $i$ is the index and $v_\lambda$ is the total number of the cells of a phase
partition. $q$ should be considered as a parameter characterizing the topological
feature of the phase space. If the phase space is a simple fractal with dimension
$d_f$, we have $q=d_f/d$ where $d$ is the dimension of the phase space when it is
smoothly occupied. Eq.(\ref{3}) is originally derived with the hypothesis of
incomplete information and incomplete probability distribution for fractal phase
space\cite{Wang04a}. But its form remains the same for complete probability
distribution $p_i$ (i.e. $\sum_{i=1}^{v}p_i=1$). The following presentation is made
within this formalism.

$R$ has the following properties
\begin{enumerate}

\item $R$ is a relative change, so $-1\leq R\leq 1$.

\item For a system composed of two sub-systems $A$ and $B$ with $q_{A}$ and $q_{B}$
satisfying the following rule of product joint probability
\begin{eqnarray}                                            \label{4}
\mu_{{i_A}{i_B}}^{q_{A+B}}=\mu_{i_A}^{q_A}\mu_{i_B}^{q_B},
\end{eqnarray}
it is easy to show the following nonadditivity :
\begin{eqnarray}                                            \label{5}
R(A+B)=R(A)+R(B)+R(A)R(B).
\end{eqnarray}

\item $R$ is positive and concave for $q>1$, and negative and convex for $q<1$. If
$q=1$, there is no fractal, so $R=0$. In other words, the fractal feature of the
phase space is responsible for the entropy change. This is in accordance with the
numerical result for some chaotic maps\cite{Gilbert}. In the case of a simple
fractal, $q=d_f/d>1$ ($q=d_f/d<1$) implies phase space expansion (contraction)
during scale refinement. So for chaotic dynamic systems, $q>1$ corresponds to
positive Lyapunov exponent (we lose knowledge about the system so that $R>0$) and
$q<1$ to negative Lyapunov exponent (we gain knowledge about the system so that
$R<0$).

\item We see that $R$ is a function of the dimension difference $d_f-d$. If we divide
$R$ by $q-1=\frac{d_f-d}{d}$, we get
\begin{eqnarray}                                            \label{c3}
H_q=\frac{R}{q-1}=-\frac{\sum_{i=1}(\mu_i-\mu_i^q)}{1-q}.
\end{eqnarray}
This is the form of Tsallis entropy\cite{Tsallis}. If $q=1$ or $d_f=d$, the phase
space is uniformly occupied and the ratio $H_1$ becomes Shannon entropy. Other
extended entropies can be obtained by defining entropy change rate in different
manners\cite{Wang04a}.

\end{enumerate}

\section{Some invariant measures for chaotic systems}
According to our method, $R$ can be extremized under the constraints associated with
the normalization $\sum_{i=1}^{v}\mu_i=1$ and with our knowledge about the random
variables $x^{\{\sigma\}}$ ($\sigma=1,2 ... w$ is the index of these variables)
concerned in the nonequilibrium process, with $w$ multipliers $\gamma_\sigma$ each
connected with an expectation
$\overline{x^{\{\sigma\}}}=\sum_{i=1}^{v}\mu_i^qx_i^{\{\sigma\}}$ where
$x_i^{\{\sigma\}}$ is the value of $x^{\{\sigma\}}$ when the system is at the state
$i$. The probability distribution is given by
\begin{eqnarray}                                            \label{7x}
\mu_i=\frac{1}{Q}(1-\sum_{\sigma=1}^{w}\gamma_\sigma x_i^{\{\sigma\}})^{1/(1-q)}.
\end{eqnarray}
This distribution can be applied to many chaotic maps. Following are some examples.

\subsection{Un model for population evolution}
The logistic map $y_{n+1}=Ay_n-By_n^2$ ($0<y_n<y_{max}$)\cite{Hilb94} is often used
for modelling the biological population evolution, where $y_n$ is the population of
the $n^{th}$ year or of $n^{th}$ order of iteration, $y_{max}=A/B$ is the maximal
population (for given living conditions) not to exceed in order that the concerned
species do not die out next year, $A$ is a positive constant connected with
population growth and $B$ is a positive constant connected with the population
decrease.

In this model, the two variables for the population evolution is $x_n$ and $x_n^2$.
So if in Eq.(\ref{7x}) one puts $x^{\{1\}}=y$ and $x^{\{2\}}=y^2$, one gets,
\begin{eqnarray}                                           \label{7xx}
\mu(x)=\frac{1}{Q}(1-\sum_{\sigma=1}^{2}\gamma_\sigma x_i^{\{\sigma\}})^{1/(1-q)}
=\frac{1}{Q}(1-\gamma_1 y-\gamma_2 y^2)^{1/(1-q)}.
\end{eqnarray}
According to the assumption of the roles of the $y$-term and $y^2$-term in the
population evolution, we can suppose that $\gamma_1$ is related to population
increase and $\gamma_2$ to population decrease. When $\gamma_1 y+\gamma_2 y^2\gg 1$,
Eq.(\ref{7xx}) becomes
\begin{eqnarray}                                           \label{7xxx}
\mu(x)=\frac{1}{Q}(-\gamma_1 y-\gamma_2 y^2)^{1/(1-q)}.
\end{eqnarray}
By comparison with the distribution given for $A=4$\cite{Hilb94} :
$\mu(x)=\frac{1}{\pi [x(1-x)]^{1/2}}$ where $x=y/y_{max}$, we obtain
$\gamma_1=-y_{max}$, $\gamma_2=1$, $Q=\pi/y_{max}$ and $q=3$. So $R>0$. This implies
an increasing entropy process with positive Lyapunov exponent .

\subsection{The continued fraction map}
The continued fraction map\cite{Beck} is a chaotic map given by
$x_{n+1}=1/x_n-\lfloor 1/x_n\rceil$ where $\lfloor 1/x_n\rceil$ is the integer part
of $1/x_n$ and $x_n$ is a real number between 0 and 1. The probability distribution
is given by $\mu(x)=1/(1+x)\ln 2$ and satisfies $\sum_{x}\mu(x)=1$. This
distribution can be obtained from Eq.(\ref{7x}) by using $w=1$, $x^{\{1\}}=x$,
$\gamma_1=-1$, $Q=\ln 2$ and $q=2$. This is also an increasing entropy process with
Lyapunov exponent $\lambda=\ln 2$\cite{Beck}.

Other application can be found in \cite{Wang04a}, e.g., for Ulam map $x_{n+1}=1-\mu
x_n^2$ ($-1<x_n<1$) with $\mu=2$, $q=3$ and for Zipf-Mandelbrot's law
$\mu(x)=\frac{A}{(1-Bx)^\tau}$, $q=1+1/\tau$.

\section{Concluding remarks}
A path information is defined in connection with different possible paths of chaotic
system moving in its phase space between two cells. On the basis of the assumption
that the paths are physically differentiated by their actions, we show that the
maximum path information leads to a path probability distribution in exponentials of
action from which the well known Fokker-Planck equation, the Fick's laws, the Ohm's
law and the Fourier's law can be easily derived in a general way. This result strongly
suggests that, for the probabilistic case of irregular dynamics, that the principle of
least action for regular dynamics should be replaced by the principle of maximization
of path information or dynamic uncertainty.

We show that an extended path information between two phase volumes can be related
to the entropy change of the processes linking the two phase volumes. Hence the
principles of least action and maximum path information suggest the maximum entropy
change which can be used to derive the most probable invariant measures for
nonequilibrium systems. This result is used to derive invariant measures for some
chaotic systems evolving in fractal phase space. We would like to emphasize that the
method of maximum entropy change is in accordance with the principle of maximum
entropy which can be used for nonequilibrium systems when available. If the system
of interest occupying the two phase volumes is in equilibrium, then maximum entropy
change of dynamic process linking the phase volumes naturally leads to the Jaynes'
principle of maximum entropy for equilibrium systems, as discussed in
\cite{Wang04xx}.

\section*{Acknowledgments}

I acknowledge with great pleasure the useful discussions with Professors A. Le
M\'ehaut\'e, F. Tsobnang, L. Nivanen, M. Pezeril and Dr. W. Li.


\begin{thebibliography}{99}

\bibitem {Hawking}
S.W. Hawking, T. Hertog, {\em Phys. Rev. D,} {\bf 66}(2002)123509;

S.W. Hawking, Gary.T. Horowitz, {\em Class.Quant.Grav.,} {\bf 13}(1996)1487

S. Weinberg, Quantum field theory, vol.II, Cambridge University Press, Cambridge,
1996 (chapter 23: extended field configurations in particle physics and treatments
of instantons)

\bibitem {Feynman}
R.P. Feynman, {\em Quantum mechanics and path integrals,\/} McGraw-Hill Publishing
Company, New York, 1965

\bibitem {Jaynes}
E.T. Jaynes, {\em The evolution of Carnot's principle}, The opening talk at the EMBO
Workshop on Maximum Entropy Methods in x-ray crystallographic and biological
macromolecule structure determination, Orsay, France, April 24-28, 1984; Reprinted
in Ericksen $\&$ Smith, Vol 1, pp. 267-282; {\em Phys. Rev.,} {\bf 106},620(1957)

\bibitem {Tribus}
M. Tribus, {\em D\'ecisions Rationelles dans l'incertain,} (Paris, Masson et Cie,
1972)P14-26; or {\em Rational, descriptions, decisions and designs,} (Pergamon Press
Inc., 1969)

\bibitem {Ruelle}
D. Ruelle, Is there a unified theory of nonequilibrium statistical mechanics?
Proceedings of the International Conference on Theoretical Physics Th-2002 (Paris,
July 22-27, 2002), Birkh\"auser Verlag, Berlin, 2004, p.489;

D. Ruelle, Extending the definition of entropy to nonequilibrium steady state, {\em
Proc. Nat. Acad. Sci.,} {\bf 100},30054(2003)

\bibitem {Cohen}
G. Gallavotti, E.G.D. Cohen, {\em Note on nonequilibrium stationary states and
entropy,} cond-mat/0312306

\bibitem {Lebowitz}
P.L. Garrido, S. Goldstein and J.L. Lebowitz, {\em The Boltzmann entropy for dense
fluid not in local equilibrium,} cond-mat/0310575

S. Goldstein and J.L. Lebowitz, {\em On the (Boltzmann) Entropy of Nonequilibrium
Systems,} cond-mat/0304251

\bibitem {Prigogine}
I. Prigogine, {\em Bull. Roy. Belg. Cl. Sci.,} {\bf 31},600(1945)

\bibitem {Paltridge}
G. Paltridge, {\em Quart. J. Roy. Meteor. Soc.,} {\bf 101},475(1975)

\bibitem {Lenton}
Tim Lenton, {\em Maximum entropy production in Edinburgh, a report of the second
'Daisyworld and beyond' workshop,} (2002), document obtained from
\url{http://www.cogs.susx.ac.uk/daisyworld/ws2002_overview.html}

\bibitem {Tsallis}
C. Tsallis, {\em J. Stat. Phys.,} {\bf 52},479(1988); C. Tsallis, F. Baldovin, R.
Cerbino and P. Pierobon, {\em Introduction to Nonextensive Statistical Mechanics and
Thermodynamics}, cond-mat/0309093

\bibitem {Wang02c}
Q.A. Wang, {\em Euro. Phys. J. B,} {\bf26}(2002)357

Q.A. Wang, {\em Chaos, Solitons $\&$ Fractals,\/} {\bf12}(2001)1431, Erranta :
cond-mat/0009343

\bibitem {Jaynes2}
E.T. Jaynes, Gibbs vs Boltzmann entropies, {\em American Journal of Physics}, {\bf
33},391(1965);

E.T. Jaynes, Where do we go from here? {\em Maximum entropy and Bayesian methods in
inverse problems}, pp.21-58, eddited by C. Ray Smith and W.T. Grandy Jr., D. Reidel
Publishing Company (1985)

\bibitem {Wang04x}
Q.A. Wang, Maximum path information and the principle of least action for chaotic
system, {\em Chaos, Solitons $\&$ Fractals,} (2004), in press; cond-mat/0405373 and
ccsd-00001549

\bibitem {Arnold}
V.L. Arnold, {\em Mathematical methods of classical mechanics,\/} Springer-Verlag,
New York, 1989

\bibitem {Kubo}
R. Kubo, M. Toda, N. Hashitsume, {\em Statistical physics II, Nonequilibrium
statistical mechanics,\/} Springer, Berlin, 1995

\bibitem {Beck}
C. Beck and F. Schl\"ogl, {\em Thermodynamics of chaotic systems}, Cambridge
University Press, 1993

\bibitem {Wang04a}
Q.A. Wang, {\em Chaos, Solitons $\&$ Fractals,} {\bf 19},639(2004)

Q. A. Wang and A. Le M\'ehaut\'e, {\em Chaos, Solitons $\&$ Fractals,} {\bf
21},893(2004)

\bibitem {Bonetto}
F. Bonetto, J.L. Lebowitz, L. Rey-Bellet, Fourier's Law: a Challenge for Theorists,
math-ph/0002052

\bibitem {Gilbert}
T. Gilbert, J.R.Dorfman, and P. Gaspard, {\em Phys. Rev. Lett.,} {\bf
85},1606(2000);

I. Claus and P. Gaspard, The fractality of the relaxation modes in deterministic
reaction-diffusion systems, cond-mat/0204264

\bibitem {Hilb94}
Robert C. Hilborn, {\em Chaos and Nonlinear Dynamics,\/} Oxford University press,
New York, 1994

\bibitem {Wang04xx}
Q.A. Wang, Action principle and Jaynes' guess method, cond-mat/0407515

\end{thebibliography}
\end{document}